\documentclass{article}
\usepackage{times}
\usepackage{amsfonts}
\usepackage{graphicx}
\usepackage[pdfmark]{hyperref}
\begin{document}
\noindent
{\Large   ON THE RICCI FLOW AND EMERGENT QUANTUM MECHANICS}
\vskip1cm
\noindent
{\bf Jos\'e M. Isidro}${}^{1}$, {\bf J.L.G. Santander}${}^{2}$ and {\bf P. Fern\'andez de C\'ordoba}${}^{3}$ \\
Grupo de Modelizaci\'on Interdisciplinar, Instituto de Matem\'atica Pura y Aplicada,\\ Universidad Polit\'ecnica de Valencia, Valencia 46022, Spain\\
${}^1${\tt joissan@mat.upv.es}, ${}^2${\tt jlgonzalez@mat.upv.es},\\ ${}^3${\tt pfernandez@mat.upv.es}

\vskip.5cm
\noindent
{\bf Abstract} The Ricci flow equation of a conformally flat Riemannian metric on a closed 2--dimensional configuration space is analysed. It turns out to be equivalent to the classical Hamilton--Jacobi equation for a point particle subject to a potential function that is proportional to the Ricci scalar curvature of configuration space. This allows one to obtain Schroedinger quantum mechanics from Perelman's action functional: the quantum--mechanical wavefunction is the exponential of $i$ times the conformal factor of the metric on configuration space. We explore links with the recently discussed {\it emergent quantum mechanics}.\\
To appear in the proceedings of DICE'08 (Castiglioncello, Italy, Sept. 2008), edited by H.-T. Elze.

\section{A conformally flat configuration space}\label{gauss}

Let us consider a 2--dimensional, closed Riemannian manifold ${M}$. In isothermal coordinates $x,y$ the metric reads
\begin{equation}
g_{ij}={\rm e}^{-f}\delta_{ij}, 
\label{metrik}
\end{equation}
where $f=f(x,y)$ is a function, hereafter referred to as {\it conformal factor}\/. Our conventions are $g=\vert\det g_{ij}\vert$ and $R_{im}=g^{-1/2}\partial_n\left(\Gamma_{im}^ng^{1/2}\right)-\partial_i\partial_m\left(\ln g^{1/2}\right)-\Gamma_{is}^r\Gamma_{mr}^s$ for the Ricci tensor, $\Gamma_{ij}^m=g^{mh}\left(\partial_ig_{jh}+\partial_jg_{hi}-\partial_hg_{ij}\right)/2$ being the Christoffel symbols. The volume element on ${M}$ equals
\begin{equation}
\sqrt{g}\,{\rm d}x{\rm d}y={\rm e}^{-f}{\rm d}x{\rm d}y. 
\label{skalar}
\end{equation}
Given an arbitrary function $\varphi(x,y)$ on ${M}$, we have the following expressions for the Laplacian $\nabla^2\varphi$ and the squared gradient $\left(\nabla\varphi\right)^2$:
\begin{equation}
\nabla^2\varphi:=\frac{1}{\sqrt{g}}\partial_m\left(\sqrt{g}g^{mn}\partial_n\varphi\right)={\rm e}^f\left(\partial_x^2\varphi+\partial_y^2\varphi\right)=:{\rm e}^fD^2\varphi,
\label{eins}
\end{equation}
\begin{equation}
\left(\nabla\varphi\right)^2:= g^{mn}\partial_m\varphi\partial_n\varphi={\rm e}^f\left[\left(\partial_x\varphi\right)^2+\left(\partial_y\varphi\right)^2
\right]=:{\rm e}^f\left(D\varphi\right)^2,
\label{drei}
\end{equation}
where $D^2\varphi$ and $\left(D\varphi\right)^2$ stand for the flat--space values of the Laplacian and the squared gradient, respectively. The Ricci tensor reads 
\begin{equation}
R_{ij}=\frac{1}{2}D^2f\,\delta_{ij}=\frac{1}{2}{\rm e}^{-f}\nabla^2f\,\delta_{ij}.
\label{ritchie}
\end{equation}
{}From here we obtain the Ricci scalar
\begin{equation}
R={\rm e}^{f}D^2f=\nabla^2f.
\label{rico}
\end{equation}
A manifold ${M}$ such as that considered here is in fact a compact Riemann surface without boundary.  Physically it will play the role of a configuration space for a certain point mechanics on ${M}$. It may be worthwhile to remember that any Riemannian metric on a 2--dimensional manifold can be written in the form (\ref{metrik}). This is often referred to as the property that any 2--dimensional metric is always conformally flat. This nice property of the 2--dimensional case no longer holds in $n\geq 3$ dimensions. We will however restrict our attention to $n=2$.

\section{A crash course in Ricci flow}\label{grisha}

{}For an introduction to the Ricci flow and its applications, a good reference is \cite{TOPPING}.  More technical are Perelman's original article and references therein \cite{PERELMAN}. Perelman's functional ${\cal F}[\varphi,g_{ij}]$ on the manifold ${M}$ is defined as 
\begin{equation}
{\cal F}[\varphi,g_{ij}]:=\int_{{M}}{\rm d}x{\rm d}y\sqrt{g}\,{\rm e}^{-\varphi}\left[\left(\nabla \varphi\right)^2+ R(g_{ij})\right],
\label{epe}
\end{equation}
where $g_{ij}$ is a metric on ${M}$ and $\varphi$ a real function on ${M}$. We will take the above equation as our starting point. It may be regarded physically as providing an action functional, on configuration space ${M}$, for the two independent fields $g_{ij}$ and $\varphi$. Now some aspects of the functional ${\cal F}[\varphi,g_{ij}]$ are worth mentioning. Setting $\varphi=0$ identically we have the Einstein--Hilbert functional for gravity on ${M}$. Admittedly Einstein--Hilbert gravity, being a boundary term in $n=2$ dimensions, is trivial in $n=2$ dimensions. However the generalisation of 2--dimensional gravity provided by the functional ${\cal F}[\varphi,g_{ij}]$ when $\varphi\neq 0$ is interesting. Indeed, Perelman's functional  arises in string theory as the low--energy effective action of the bosonic string \cite{STRING}.  Already these two properties suffice to justify our interest in the functional (\ref{epe}). If this were not enough, the mathematical applications of Ricci--flow theory are impressive \cite{PERELMAN, TOPPING, CARROLL9}, although they will not be dealt with here. Instead we will concentrate our attention on the relation between the Ricci flow and quantum mechanics on ${M}$.

We first compute the Euler--Lagrange extremals corresponding to the fields $g_{ij}$ and $\varphi$. Next we set the equations of motion so obtained equal to the first--order time derivatives of  $g_{ij}$ and $\varphi$, respectively. This results in the two evolution equations 
\begin{equation}
\frac{\partial g_{ij}}{\partial t}=-2\left(R_{ij}+\nabla_i\nabla_j\varphi\right), \quad \frac{\partial \varphi}{\partial t}=-\nabla^2\varphi-R.
\label{evolut}
\end{equation}
We stress that the right--hand sides of (\ref{evolut}), once equated to zero, are the Euler--Lagrange equations of motion corresponding to (\ref{epe}), and that the time derivatives on the left--hand sides have been put in by hand. In fact time is not a coordinate on configuration space ${M}$, but an external parameter. The two equations (\ref{evolut}) are referred to as the {\it gradient flow}\/ of ${\cal F}$, since they provide a set of time--evolution equations. Via a time--dependent diffeomorphism, one can show that the set (\ref{evolut}) are equivalent to
\begin{equation}
\frac{\partial g_{ij}}{\partial t}=-2R_{ij}, \quad \frac{\partial \varphi}{\partial t}=-\nabla^2\varphi+\left(\nabla\varphi\right)^2-R.
\label{voluzione}
\end{equation}
We will use (\ref{voluzione}) rather than (\ref{evolut}).

As already remarked, one advantage of having a 2--dimensional configuration space is that all metrics on it are conformal, so we can substitute (\ref{metrik}) throughout. By (\ref{skalar}) and (\ref{rico}), we can express ${\cal F}[\varphi, g_{ij}]$ as 
\begin{equation}
{\cal F}[\varphi, f]:={\cal F}[\varphi, g_{ij}(f)]=\int_{{M}}{\rm d}x{\rm d}y\,{\rm e}^{-\varphi-f}\left[\left(\nabla\varphi\right)^2+\nabla^2f\right].
\label{epep}
\end{equation}
In order to understand the physical meaning of the flow eqns. (\ref{voluzione}), let us analyse them in more detail. Using (\ref{metrik}) and (\ref{ritchie}) we see that the first flow equation,
\begin{equation}
\frac{\partial g_{ij}}{\partial t}= -2R_{ij},
\label{bella}
\end{equation}
 is equivalent to 
\begin{equation}
\frac{\partial f}{\partial t}=\nabla^2f.
\label{chaleur}
\end{equation}
This is the usual heat equation, with the important difference that the Laplacian operator $\nabla^2$ is given by (\ref{eins}):  indeed $M$ is not flat but only {\it conformally}\/ flat. So conformal metrics on the (curved) manifold ${M}$ evolve in time according to the heat equation wih respect to the corresponding (curved) Laplacian. The second flow equation in (\ref{voluzione}) will be the subject of our attention in what follows.

So far, the conformal factor $f$ and the scalar $\varphi$ have been considered as independent fields. Setting now $\varphi=f$ in (\ref{epep}) we obtain 
\begin{equation}
{\cal F}[f]:={\cal F}[\varphi =f,f]=\int_{{M}}{\rm d}x{\rm d}y\,{\rm e}^{-2f}\left[\left(\nabla f\right)^2+\nabla^2f\right].
\label{tonton}
\end{equation}
After setting $\varphi=f$ we appear to have a contradiction, since we have two different flow equations in  (\ref{voluzione}) for just one field $f$.
That there is in fact no contradiction can be seen as follows. In (\ref{voluzione}) we have two independent flow equations for the two independent fields $f$ and $\varphi$. Equating the latter two fields implies that the two flow equations must reduce to just one. This can be achieved by substituting one of the two flow eqns. (\ref{voluzione}) into the remaining one. By (\ref{rico}) and (\ref{chaleur}) we have $R=\partial f/\partial t$, which substituted into the second flow equation of (\ref{voluzione}) leads to
\begin{equation}
\frac{\partial f}{\partial t}=\frac{1}{2}\left(\nabla f\right)^2-\frac{1}{2}\nabla^2f.
\label{tott}
\end{equation}
We will later on find it useful to distinguish notationally between the time--independent conformal factor $f$, as it stands in the functional (\ref{tonton}), and the time--dependent conformal factor as it stands in the flow equation (\ref{tott}). We therefore rewrite (\ref{tott}) as
\begin{equation}
\frac{\partial \tilde f}{\partial t}=\frac{1}{2}\left(\nabla \tilde f\right)^2-\frac{1}{2}\nabla^2\tilde f,
\label{konvfluss}
\end{equation}
where a tilde on top of a field indicates that it is a time--dependent quantity.

\section{A correspondence between conformal metrics and mechanical actions}\label{buch}

In what follows we will regard the manifold ${M}$ as the configuration space of a mechanical system, to be identified presently.  We will establish a 1--to--1 correspondence between conformally flat metrics on configuration space ${M}$, and (classical or quantum) mechanical systems on ${M}$. Let us consider classical mechanics in the first place. We recall that, for a point particle of mass $m$ subject to a time--independent potential $U$, the Hamilton--Jacobi equation for the time--dependent action $\tilde S$ reads
\begin{equation}
\frac{\partial \tilde S}{\partial t}+\frac{1}{2m}\left(\nabla \tilde S\right)^2+U=0.
\label{hamjacbtrev}
\end{equation}
It is well known that, separating the time variable as per 
\begin{equation}
\tilde S=S-Et,
\label{trennung}
\end{equation}
with $S$ the so--called reduced action, one obtains
\begin{equation}
\frac{1}{2m}\left(\nabla S\right)^2+U=E.
\label{stillnight}
\end{equation}
Eqn. (\ref{trennung}) suggests separating variables in (\ref{konvfluss}) as per
\begin{equation}
\tilde f=f+Et,
\label{getrennt}
\end{equation}
where the sign of the time variable is reversed\footnote{This time reversal is imposed on us by the time--flow eqn. (\ref{konvfluss}), with respect to which time is reversed in the mechanical model. This is just a rewording of (part of) section 6.4 of ref. \cite{TOPPING}, where a corresponding heat flow is run {\it backwards}\/ in time.}
 with respect to (\ref{trennung}). Substituting (\ref{getrennt}) into  (\ref{konvfluss}) leads to 
\begin{equation}
\frac{1}{2}\left(\nabla f\right)^2-\frac{1}{2}\nabla^2f=E.
\label{masymas}
\end{equation}
Comparing (\ref{masymas}) with  (\ref{stillnight}) we conclude that, picking a value of the mass $m=1$, the following identifications can be made:
\begin{equation}
S=f, \qquad U=-\frac{1}{2}\nabla^2f = -\frac{1}{2}R.
\label{geoint}
\end{equation}
So the potential $U$ is proportional to the scalar Ricci curvature of the configuration space  ${M}$, while the reduced action $S$ equals the conformal factor $f$. This concludes the first half of our dictionary: to construct a classical mechanics starting from a conformal metric on ${M}$. 

Conversely, if we are given a classical mechanics as determined by an arbitrary potential function $U$ on ${M}$, and we are required to construct a conformal metric on ${M}$, then the solution is given by the function $f$ satisfying the Poisson equation $-2U=\nabla^2f$, where the Laplacian is computed with respect to the unknown function $f$.

Although we have so far considered the {\it classical}\/ mechanics associated with a given conformal factor, one can immediately construct the corresponding {\it quantum}\/ mechanics, by means of the Schroedinger equation for the potential $U$.  We can therefore restate our result as follows: we have established a 1--to--1 correspondence between conformally flat metrics on configuration space, and quantum--mechanical systems on that same space.

\section{Schroedinger's functional from Perelman's functional}\label{diskk}

Let us summarise our results. We have considered a conformally flat Riemannian metric on a closed  2--dimensional manifold ${M}$, and regarded the latter as the configuration space of a classical mechanical system. We have formulated a dictionary between such conformal metrics, on the one hand, and quantum mechanics on the same space, on the other. This dictionary has a nice geometrical interpretation: the reduced mechanical action $S$ equals the conformal factor $f$, and the potential function $U$ is (proportional to) the Ricci curvature of ${M}$. It is interesting to observe that the Ricci scalar as a potential function has also arisen in the context of Bohmian mechanics \cite{KOCH1, KOCH2, KOCH3}.

The previous correspondence can be exploited further: we will exchange a conformally flat metric for a wavefunction satisfying the Schroedinger equation for the potential $U$.  We first observe that the Schroedinger equation itself can be obtained as the extremal of the action functional
\begin{equation}
{\cal S}[\psi,\psi^*]:=\int_{{M}}{\rm d}x{\rm d}y\sqrt{g}\left({\rm i}\psi^*\frac{\partial\psi}{\partial t}-\frac{1}{2m}\nabla\psi^*\nabla\psi-U\psi^*\psi\right).
\label{actwelle}
\end{equation}
Pick  $m=1$ as before, and substitute 
\begin{equation}
\psi={\rm e}^{{\rm i}\tilde f}
\label{wellenfunktion}
\end{equation}
into (\ref{actwelle}) to obtain  $-\partial_t\tilde f-\frac{1}{2}(\nabla \tilde f)^2+\frac{1}{2}\nabla^2\tilde f$ within the integrand. As before, let us consider the stationary case, where $\partial_t \tilde f=0$ and $\tilde f$ becomes $f$. Then (\ref{actwelle}) turns into
\begin{equation}
{\cal S}[f]:={\cal S}\left[\psi={\rm e}^{{\rm i} f}\right]=\frac{1}{2}\int_{{M}}{\rm d}x{\rm d}y\,{\rm e}^{-f}\left[-(\nabla f)^2+\nabla^2f\right].
\label{esseprima}
\end{equation}
Comparing the functionals (\ref{tonton}) and (\ref{esseprima}) we arrive at the following interesting relation:
\begin{equation}
{\cal F}\left[f/2\right]={\cal S}[f]+\frac{3}{2}{\cal K}[f],
\label{irreplaceble}
\end{equation}
where 
\begin{equation}
{\cal K}[f]:=\frac{1}{2}\int_{{M}}{\rm d}x{\rm d}y\,{\rm e}^{-f}\left(\nabla f\right)^2
\label{kinetic}
\end{equation}
is the kinetic energy functional on $M$. In ${\cal F}\left[f/2\right]$ above, the Laplacian $\nabla^2f$ and the squared gradient $(\nabla f)^2$ are computed with respect to the conformal factor $f$, even if the functional ${\cal F}$ is evaluated at $f/2$. Thus, on a compact Riemann surface without boundary, the Schroedinger functional ${\cal S}[f]$ turns out to be a close cousin of the Perelman functional ${\cal F}[f/2]$.  Altogether we have proved that Schroedinger quantum mechanics on a 2--dimensional, compact configuration space arises from Perelman's functional.

\section{Discussion and perspectives}\label{intt}

The Ricci flow has provided many far--reaching insights into long--standing problems in topology and geometry \cite{PERELMAN, TOPPING}. Recent works \cite{MATONE1, MATONE2, CARROLL1, CARROLL2, CARROLL5, CARROLL6, CARROLL7, CARROLL8} have shed light on applications of conformal symmetry and the Ricci flow to foundational issues in quantum mechanics. In this contribution we have established a 1--to--1 correspondence between conformally flat metrics on configuration space, and quantum mechanics on that same space. This correspondence has been used to prove that Schroedinger quantum mechanics in two space dimensions arises from Perelman's functional on a compact Riemann surface. 

We have worked in the 2--dimensional case for simplicity. Now Perelman's functional  arises in string theory as the low--energy effective action of the bosonic string \cite{STRING}. In view of these facts it is very tempting to try and interpret quantum mechanics itself, in any number of dimensions,  possibly also noncompact, as some kind of effective, low--energy approximation to some more fundamental theory. Related ideas have been put forward in the literature \cite{THOOFT1, THOOFT2, ELZE1, ELZE5}, where standard quantum mechanics has been argued to {\it emerge}\/ from an underlying deterministic theory.  Basically, in {\it emergent}\/ quantum mechanics, one starts from a {\it deterministic}\/ model and, via some dissipative mechanism that implements information loss, one ends up with a {\it probabilistic}\/ theory. Several mechanisms implementing information loss have been proposed in the literature. Thus, in ref. \cite{THOOFT1}, dissipation is effected by an attractor on phase space, which produces a lock--in of classical trajectories around some fixed point; instead, in ref. \cite{ELZE5} dissipation arises as a coarse--graining of classical information via a probability distribution function on phase space. A somewhat different dissipative mechanism, based on the Ricci flow equation (\ref{bella}), has been put forward in ref. \cite{ISIDRO}.

Some features of emergent quantum mechanics are present in our picture. Most notable among them is the presence of dissipation, or information loss: as remarked above, this underlies the passage from a classical description to a quantum description. Indeed, in our setup, the classical description is provided by the conformal factor $f$ of the metric $g_{ij}$ on configuration space, while the quantum description is given by the wavefunction $\psi={\rm e}^{{\rm i}f}$. The latter contains less information than the former, as there exist different conformal factors $f$ giving rise to just one quantum wavefunction $\psi$. This situation is analogous to that described in \cite{THOOFT1, THOOFT2}, in which quantum states arise as equivalence classes of classical states: different classical states may fall into one and the same quantum equivalence class. Beyond the trivial case of any two conformal factors $f_1$ and $f_2$ differing by $2\pi$ times an integer, there is the more interesting case of $f_1$ and $f_2$ satisfying $\nabla^2_1f_1=\nabla^2_2f_2$, where the subindices $1,2$ refer to the fact that the corresponding Laplacians are computed with respect to the conformal factors $f_1$ and $f_2$, respectively. If $M$ is such that the Laplace--like equation $\nabla^2_1f_1-\nabla^2_2f_2=0$ admits nontrivial solutions, then any two such $f_1$ and $f_2$ (different classical states) fall into the same quantum state, as both $f_1$ and $f_2$ give rise to the same potential function $-2U=\nabla^2_1f_1=\nabla^2_2f_2$. 

Another feature of emergent quantum mechanics that is present in our picture is the following. In refs.  \cite{THOOFT1, THOOFT2, ELZE1} it has been established that to every quantum system there corresponds at least one deterministic system which, upon prequantisation, gives back the original quantum system. In our setup this existence theorem is realised alternatively as follows. Let a quantum system possessing the potential function $V$ be given on the configuration space $M$, the latter satisfying the same requirements as above. Let us consider the Poisson equation on $M$, $\nabla^2_Vf_V=-2V$, where $f_V$ is some unknown conformal factor, to be determined as the solution to this Poisson equation, and $\nabla^2_V$ is the corresponding Laplacian. We claim that the deterministic system, the prequantisation of which gives back the original quantum system with the potential function $V$, is described by the following data: configuration space $M$, with classical states being conformal factors $f_V$ and mechanics described by the action functional (\ref{tonton}). The lock--in mechanism (in the terminology of refs. \cite{THOOFT1, THOOFT2}) is the choice of one particular conformal factor, with respect to which the Laplacian is computed, out of all possible solutions to the Poisson equation on $M$, $\nabla^2_Vf_V=-2V$. The problem thus becomes topological--geometrical in nature, as the lock--in mechanism has been translated into a problem concerning the geometry and topology of configuration space $M$, namely, whether or not the Poisson equation possesses solutions on $M$, and how many.
 
As a perspective for future work, it would be an interesting question to ask whether or not Ricci--flow techniques could  be used to implement some kind of renormalisation--group flow from an underlying deterministic model to an emergent quantum mechanics. This would be in line with the view that quantum mechanics is an infrared phenomenon \cite{ELZE10, ELZE11}, and also with models of spacetime whereby classical gravity has been argued to arise by some process of {\it thermalisation}\/ of some underlying quantum theory. We hope to address these issues in the future.

\vskip.5cm
\noindent
{\bf Acknowledgements} {It is a great pleasure to thank H.-T. Elze for interesting technical discussions. J.M.I. also thanks Max--Planck--Institut f\"ur Gravitationsphysik, Albert--Einstein--Institut (Potsdam, Germany) for hospitality extended over a long period of time.}

\end{document}